# Low-threshold, highly stable colloidal quantum dot short-wave infrared laser enabled by suppression of trap-assisted Auger recombination

Nima Taghipour, Guy L. Whitworth, Andreas Othonos, Mariona Dalmases, Santanu Pradhan, Yongjie Wang, Gaurav Kumar and Gerasimos Konstantatos*

N. T, Dr. G. W, Dr. M. D, Dr. S. P, Y. W, G. K, Prof. G. K.
ICFO, Institut de Ciències Fotòniques, The Barcelona Institute of Science and Technology, 08860 Barcelona, Spain
E-mail: Gerasimos.konstantatos@icfo.eu
Prof. A. O.
Laboratory of Ultrafast Science, Department of Physics, University of Cyprus, Nicosia 1678, Cyprus
Prof. G. K.
ICREA—Institució Catalana de Recerca i Estudis Avançats, Barcelona, Spain

**Pb-chalcogenide colloidal quantum dots (CQDs) are attractive materials to be used as tuneable laser media across the infrared spectrum. However, excessive nonradiative Auger recombination due to the presence of trap states outcompetes light amplification by rapidly annihilating the exciton population, leading to high gain thresholds. Here, we employ a binary blend of CQDs and ZnO nanocrystals in order to passivate the in-gap trap states of PbS-CQD gain medium. Using transient absorption, we measure a five-fold increase in Auger lifetime demonstrating the suppression of trap-assisted Auger recombination. By doing so, we achieve a two-fold reduction in amplified spontaneous emission (ASE) threshold. Finally, by integrating our proposed binary blend to a DFB resonator, we demonstrate single-mode lasing emission at 1650 nm with a linewidth of 1.23 nm (0.62 meV), operating at a low lasing threshold of ~385 $\mu J.cm^{-2}$. The Auger suppression in this system has allowed to achieve unprecedented lasing emission stability for a CQD laser with recorded continuous operation of 5 hours at room temperature and ambient conditions.**

**Introduction**

Semiconductor colloidal quantum dots (CQDs) have shown great promise as gain materials for solution-processed lasers, ranging from the visible[1–5] to the infrared.[6,7] However, light amplification in CQDs is severely hindered by intrinsic multiexitonic interactions, stemming from non-unity degeneracy of band-edge states.[1,6] This results in high lasing thresholds in QDs, as well as unfavourable nonradiative Auger recombination,[1–3,8,9] whereby the exciton energy is nonradiatively transferred to a third carrier. Recent advances in the synthesis of Cd-chalcogenides CQDs, –particularly the core-giant shell structures, has led to drastically suppressed Auger recombination for the realization of efficient lasing performance in the visible.[2–5,9–15] Recently, room temperature infrared ASE[6] and lasing[7] in undoped and heavily n-doped PbS QD films have been successfully demonstrated. Yet, those were achieved under high optical pump intensities, where the ASE and lasing thresholds for neutral PbS QD films is in the range of 800-1500 $\mu J.cm^{-2}$.[6] Moreover, the reported biexciton Auger lifetimes for PbS QDs film is ~200 ps,[6] which is an order of magnitude faster than engineered CdSe-based QDs.[9,11,15] The high ASE and lasing thresholds in PbS QD films are attributed to very fast Auger process and high degeneracy of Pb-chalcogenide QDs (8-fold).

The photoexcited carriers in QDs can be captured by mid-gap surface trap-states, resulting in long-lived net charges. The passivation of these surface trap states has been demonstrated to improve performance in CQDs solar cells[16–19] and light emitting diodes (LEDs),[20,21] namely optoelectronic devices operating in the low carrier density regime (carrier per dot <<1). The role of mid-gap trap states on the performance of optoelectronic devices operating in high carrier density mode, such as lasers, has thus far remained underexplored. The presence of mid-gap trap states, however, is expected to hamper efficient light amplification in CQDs, due to a nonradiative recombination mechanism that onsets in multi-carrier conditions, known as trap-assisted Auger recombination.[22–25] Here, we posit that by reducing the trap-state density in PbS QDs, induced by an appropriate matrix[15,19] would enable us to suppress the trap-assisted Auger process leading to low optical gain and lasing thresholds. To do so, we have considered a thin film architecture made up of a binary blend of PbS-emitter QDs and large band-gap n-type ZnO nanocrystals (NCs). The proposed binary blend enabled us to realize a record low-threshold and highly stable single-mode lasing emission in the infrared owing to suppression of the Auger recombination and reduction of self-absorption losses in the gain medium. To our knowledge this constitutes the first demonstration of suppressing Auger recombination in CQD solids by engineering at the supra-nanocrystalline level.

**Engineering of Auger decay at supra-nanocrystalline level**

**Figure 1** schematically shows the concept of suppression of trap-assisted Auger processes. Common Auger recombination (**Figure 1a**), the so-called trimolecular process, is an intrinsic material process which depends on the level of the excitation regardless of the number of trap states. **Figure 1b** illustrates another type of Auger process, which is mediated by the mid-gap trap states and can be distinguished by its bimolecular kinetics.[24–26] These bimolecular Auger processes dominate the non-radiative recombination in the saturation pump fluence regime where the number of photogenerated charge-carriers, $\Delta n$, is larger than the concentration of trap states, $n_1, p_1$ ( $\Delta n \gg n_1, p_1$[22,24]). At the operating fluences needed to observe optical gain in PbS-emitter QDs, the bimolecular Auger is the dominant nonradiative process. Therefore, to obtain an efficient infrared laser based on PbS QDs, it is essential to supress the trap-assisted Auger decay. In our proposed binary blend architecture, by bringing ZnO NCs in the vicinity of PbS- emitter QDs, the deep electron acceptor sites of PbS QDs are passivated by the transferred electrons from electron-rich ZnO NCs.[16,20] (see **Figure 1c**)

To validate our hypothesis about the trap-state passivation, we conducted thermal admittance spectroscopy (TAS) (See Methods). TAS enables us to characterize the energy and number density of trap states in photosensitive materials.[17,27] The results and detailed analysis of our TAS are provided in Supplementary Note 1. The distribution of the trap-state density is plotted as a function of trap-state position ($E_T$) in Supplementary **Figure S3**. There is a clear signature of the trap-state reduction using QDs/NCs blending, where $E_T$ is decreased from 0.1 eV in the bare PbS QD device to 0.046 eV in the PbS/ZnO blend. Consequently, the overall trap-state density dropped by an order of magnitude, from 1.35× $10^{17}$ in bare-PbS CQD device to 2.36 × $10^{16}$ in PbS/ZnO binary blend with 30% ZnO content. Therefore, with the reduction in the trap-state density by one order of magnitude, the probability of electron capturing should be significantly reduced in PbS-emitter QDs which plays a central role in trap-assisted Auger recombination.

**Single-exciton and multiexciton dynamics**

We investigated the effect of passivating trap-states in our binary CQD films on excitonic kinetics. To do so, we performed time-resolved photoluminescence (PL) and PL quantum yield (PLQY) measurements to understand the single-exciton kinetics (low injection regime). The details of time-resolved PL and PL-QY measurements are described in Methods Section. **Figure 2a** shows the single-exciton dynamics of bare-PbS and PbS/ZnO binary blend samples, where the fitting lifetime components are clearly prolonged in the blend (fitting values are given in Supplementary **Table. 1**). This can be explained by the successful remote passivation of the electron trap-states[20] in PbS-emitter QDs. Following this, PLQY is drastically increased to ≈30-35% in the PbS/ZnO binary blend film, while ≈2-3% for the bare-PbS emitter sample. Therefore, the nonradiative recombination associated to the trap-states is strongly suppressed in the single-exciton regime.

Next, to evaluate the multiexciton behaviour of bare-PbS emitter QDs and PbS/ZnO binary blend samples, we performed femtosecond transient absorption (TA) spectroscopy (See Methods Section). The multiexciton dynamics of the samples were extracted using the simple subtractive method described by Klimov et al[8] (See Supplementary **Note 3**). To extract the bi-exciton Auger decay in bare-PbS and PbS/ZnO binary blend samples, the recorded trace at low fluences (⟨N⟩ ≪ 1, where ⟨N⟩ is the average number of excitons per dot) was subtracted from the traces at $\langle N \rangle \approx 0.4$. Note that at these values of $\langle N \rangle$, the excess carrier density is on the order of $\Delta n \propto 10^{19}$ where trap-assisted Auger processes are dominant. (See Supplementary **Note 3.3** for estimation of $\Delta n$). As shown in **Figure 2b**, the bi-exciton lifetimes are prolonged by increasing the loading amount of ZnO in binary samples for ~5.8 nm PbS QDs. This trend is also observed for different sizes of PbS-emitter QDs, $D_{CQDs}$= 6.2 nm (**Figure 2c**) and $D_{CQDs}$= 5.6 nm (**Figure 2d**). For the case of $D_{CQDs}$= 5.6 nm (**Figure 2d**), the bi-exciton lifetime is extended up to 1000 ps in the PbS/ZnO blend, which is 5 times longer than that of pure PbS. The obtained lifetime is the longest reported bi-exciton Auger recombination lifetime in PbS(Se)-based QDs, which give optical gain. Moreover, as shown in **Figure 2e** and **Figure 2f**, the 3-exciton and 4-exciton Auger process are also prolonged in the binary blend architectures. The multiexcitonic Auger lifetime in PbS QDs can be estimated as $\tau_{N_e,N_h,A} = 8\tau_{2A}/[(N_e)N_h(N_e + N_h - 2)]$, where $N_e = N_h$ are the number of excited electron and holes, $\tau_{2A}$ is the biexciton Auger lifetime. Then, this expression suggests a ratio of $\tau_4:\tau_3:\tau_2 =$ 0.08:0.22:1 which is in good agreement with the ratio 0.085:0.19:1 of experimentally obtained lifetimes for 4-, 3- and 2- excitons relaxation of 5.6 nm PbS QDs(See Figure 2d-f). It is further noteworthy that the standard Auger size scaling law ($1/R^3$) is not evident in our samples. This is expected by the origin of the Auger process studied herein of close-packed ligand exchanged solid state films. In this setting Auger is not solely governed by the confinement of the carriers, that gives rise to Auger scaling reported in isolated quantum dots in solution. Instead Auger processes in our films is determined by the surface of the dots, the ligand exchange process and the local environment, factors that do not follow a size dependence.

**ASE and lasing characterizations**

To investigate the effect of supressing trap-assisted Auger processes on optical gain performance, we carried-out ASE measurements in bare-PbS QDs and Pb/ZnO binary blend samples. The details of the ASE measurements can be found in Methods Section. As an exemplary case, the ASE spectra along with the full-width at half-maximum (FWHM) of a PbS/ZnO binary blend with 30% ZnO loading are plotted in **Figure 3a** as a function of absorbed fluence (See Supplementary **Figure S6-8** for other sizes). At low fluences, only spontaneous emission observed in PL spectra within a FWHM of ~94 nm (**Figure 3a**). With increasing pump intensity, the ASE peak is evident at ~1605 nm in the emission spectra, having a narrower FWHM of ~14 nm. In **Figure 3b**, we plotted the integrated intensity emission as a function of the absorbed fluence at the optical pumping wavelength (1030 nm) for bare-PbS QDs and varying ZnO loading in PbS/ZnO binary blend. By increasing the pump intensity, the integrated PL intensity increases linearly at low fluences, and then at ASE threshold exhibits a sharp transition to super-linear behaviour (**Figure 3b**). The ASE threshold (of absorbed pump fluence) for bare-PbS QDs is 138 $\mu$J.cm$^{-2}$, while for 20%, 25% and 30% ZnO loading in PbS/ZnO binary blends is 80, 58 and 39 $\mu$J.cm$^{-2}$, respectively; substantially reducing the ASE threshold up to 3-fold. We attribute this substantial reduction in ASE thresholds to: Firstly, the significant suppression of Auger recombination in our binary films, which limits the ASE and lasing thresholds in QD systems[15] and secondly, to reduced self-absorption losses at ASE wavelength in binary blends due to the decreased number of PbS-emitter QDs per unit volume of the gain medium, which has been experimentally shown to be beneficial for reducing the ASE and lasing thresholds in other types of engineered QDs.[2,9,10] Herein, we reduced the self-absorption losses by replacing the gain material (i.e., PbS-emitter QD) with a non-absorbing material (i.e., ZnO NC) which is more robust method compared to the methods previously reported.

In addition, varying the ZnO loading in binary blend led us to achieve fine-tuneable gain spectrum across the optical telecommunication band as depicted in **Figure 3c**, with resolution beyond what quantum size effect can achieve. Here, by increasing ZnO content in blend samples, the ASE spectra is gradually blue-shifted in the examined three different sizes of PbS-emitter QDs. For instance, in 5.6 nm QDs, the ASE spectra of PbS/ZnO blend samples having 15%, 20% and 25% ZnO loading is correspondingly blue-shifted by ~17, 24, 30 nm with respect to their bare-PbS counterpart (light-orange coloured spectra in **Figure 3c**, A similar blue-shift is observed in the absorption spectra (See Supplementary **Figure S9**). We ascribe the blue-shifting of ASE and absorption spectra due to increasing inter-dot spacing of PbS-emitter CQDs[28,29], **Figure 3d** provides a summary of the measured ASE in different sizes of PbS-emitter QDs as a function of ZnO content. As shown in **Figure 3d**, in all sizes of QDs, the ASE thresholds are clearly reduced by increasing the amount of ZnO loading in binary blend. For instance, the pump fluence ASE threshold for bare-PbS QDs having a diameter size of 5.8 nm is ~1060 $\mu J.cm^{-2}$, which drops to ~430 $\mu J.cm^{-2}$ in the binary blend sample with 30% ZnO loading.

Finally, we combined the PbS/ZnO binary blend with 30% ZnO loading (our best performing ASE sample) with a second-order distributed feedback (DFB) resonator which provides strong in-plane feedback and lasing emission normal to the surface. **Figure 4a** illustrates the schematic of the DFB laser, where the film is pumped by a stripe excitation at normal incidence while the lasing emission is emitted from the surface of the device. A cross-sectional scanning electron microscopy (SEM) image of the PbS/ZnO binary blend on top of the DFB grating is given in **Figure 4b**, along with its parameters. By increasing the pump fluence, a single-mode lasing peak emerges at ~1650 nm with a linewidth of 1.26 nm (0.62 meV) (**Figure 4c**). This is accompanied by a clearly sharp growth in emission intensity at a pump fluence of ~385 $\mu J.cm^{-2}$ (corresponding to ~46 $\mu J.cm^{-2}$ in terms of absorbed fluence) indicating lasing threshold (**Figure 4d**). The lasing emission intensity follows a S-shaped behaviour (changing the slopes in log-log scale) as a function of the pump intensity, which is a characteristic of lasing action (**Figure 4d**). This is the record lowest lasing threshold among other PbS-based QDs infrared lasers.[7] In addition, our DFB laser devices show a lasing emission with an average FWHM of 1.45± 0.21 nm (0.68± 0.09 meV) (See Supplementary **Figure S10-12** for other DFB lasers) outperforming the previously reported PbS-based DFB lasers.[7] This can be attributed to the reduction of the refractive index contrast between the gain medium (i.e., PbS/ZnO (30%) blend) and the grating (i.e., $Al_2O_3$),[7] and the reduction of self-absorption in the cavity. **Figure 4f** plots the lasing emission intensity versus the angle of the polarizer between the device and detector. The experimental data (symbols) are well-fitted with a function of $cos^2(\theta)$, indicating a linearly polarized lasing emission with a polarization factor of R= 0.96, R is defined as $(I_\parallel - I_\perp)/(I_\parallel + I_\perp)$, where $I_\parallel$ and $I_\perp$ is the emission intensity parallel and vertical to the optical axis, respectively.[10] We further studied the lasing stability of our devices under continuous excitation of a pulsed-laser. As shown in **Figure 4g**, the lasing intensity is dropped to only 15% of the initial value after 5 hours continuous excitation at 10 kHz (corresponds to 180 millions laser shots). Notably the obtained stability in the present study is the highest lasing stability reported to date among all CQD-based lasers. We ascribe this to the significant reduction of Auger-assisted recombination, which cause heating issues in QDs thin films leading to material degradation. The reduction in ASE and lasing thresholds owing to suppression of the trap-assisted Auger recombination in PbS/ZnO binary blend is consistently witnessed in all studied devices with an average reduction of 1.8-2.4 (See **Figure 3d**, **Figure 4d** and Supplementary **Figure S6-8**).

**Conclusion**

Previously the use of doping had been reported as a means to reduce gain threshold by populating the first excited state of PbS CQDs allowing to reach population inversion at lower pump fulence[6,7]. In the present study we sought to supress trap-assisted Auger, a competing to gain mechanism that has been know to increase ASE and lasing thresholds in gain media in excess of their theoretical values. In doing so, we report a new approach to improve the optical gain performance in QDs systems by remote passivation of the in-gap trap states with the use of an appropriate mixture of CQDs with ZnO NCs. While doping and Auger supression have led to lasing threshold reduction the underlying mechanims are different, motivating further work in the filed to synergistically combine the two techniques for further threshold reduction.

Last but not least, this method led us to achieve a record stable lasing performance as well as an ultra-fine tuneable gain spectrum within the optical communication window.

**Experimental Section**

*PbS QDs synthesis:*

PbS QDs were synthesized by a multiple injection method, under inert atmosphere. Briefly, for 5.6 nm PbS QDs, 0.446g PbO, 3.8mL oleic acid and 50 mL 1-octadecene (ODE) were pumped for 1h at 100°C to form lead oleate. Once under argon, the solution was kept at 100°C and a solution of 70 µL hexamethyldisilathiane (HMS) in 3 mL ODE was quickly injected. After 6 minutes of reaction, a second solution of 90 µL HMS in 9 mL ODE was dropwise injected. Afterwards, the solution was cooled down naturally to room temperature and PbS QDs were precipitated with the addition of a mixture of acetone/ethanol, redispersing in anhydrous toluene. This process was repeated two more times for further purification. Finally, the concentration of PbS QDs was adjusted to 30 mg/mL and the solution was stored in the glovebox to avoid oxidation. NCs size was tuned by the variation of the amount of HMS used in each solution. For 5.8 nm QDs 65 µL HMS was used in the first injection, and 75 µL in the second. For 6.2 nm QDs, the amount of HMS added in the first injection was 60 µL and in the second, 80µL.

.

***ZnO NCs synthesis:*** ZnO NCs were synthesized based on the method shown previously[16]. Zinc acetate dihydrate (2.95 g) was dissolved in methanol (125 mL) on a 250mL flask, under continuous stirring, and the temperature was set at 60°C. Simultaneously, another solution was prepared by dissolving 1.48 g of KOH in 65 mL of methanol. The prepared solution of KOH was slowly (in 4 min) added to the zinc acetate solution and the reaction was left unchanged for 2.5 h. After completion of the reaction, the heating was removed, and the solution was led to slowly cool down to room temperature under constant stirring. Prepared ZnO NCs solution was centrifuged at 3500 r.p.m. for 5 min. Then, the supernatant was discarted and the remaining NCs were dispersed in methanol, this centrifuge process was repeated three times. Finally, the ZnO NCs was dispersed in a solution of 5% butylamine in toluene for making the binary blend solutions, and in 2% butylamine in chloroform for formation of the base layer.

***Active layer deposition:*** Active layer was formed on the desired substrate by spin coating technique. First, the substrate was covered with 50 µL of QDs solution and spin coating started after 5 s, within a spin speed of 2500 r.p.m. for 15 s. Then, a mixed ligand of $ZnI_2$/MPA (mixing of 25 mM zinc iodide ($ZnI_2$) in methanol and 0.015% 3-mercaptopropionic acid (MPA) in methanol) was added on the formed surface. The spin coating was started after 7 s, at a spin speed of 2500 r.p.m. for 60 s, while a few drops of methanol were poured in order to clean the film from any unwanted organics. These steps were repeated until reaching the desired thickness. For bare-PbS QDs solutions, the QDs solution was prepared within a concentration of 30 mg. $mL^{-1}$. Before mixing, the ZnO NCs were prepared in separate vials with the same concentration (30 mg. $mL^{-1}$). For PbS/ZnO binary blend solutions, the PbS QDs solution was mixed with ZnO solution at different volume ratios.

***Photovoltaic device (PV) preparation:*** The PV devices were prepared on a pre-cleaned indium tin oxide (ITO) coated glass substrate. The electron-transporting layer (ZnO) was prepared by spin coating of ZnO NCs in chloroform (40 mg. $mL^{-1}$) on the substrate at a speed of 4000 r.p.m for 60 s. The thickness of this layer is about 40 nm. Then, the active layer of bare-PbS or PbS/ZnO binary blend was deposited on this layer. The active layer deposition follows the similar procedure as described above. In this case, the thickness of the active layer was ~200-220 nm. Next, two layers of ethanedithiol-treated PbS (~30–35 nm) were deposited on the active layer solid film, acting as the electron-blocking layer. Finally, the back-electrode was formed by deposition of approximately 100 nm Au through a pre-patterned mask in thermal evaporator (Nano 36 Kurt J. Lesker). The active area of each device is 3.14 mm$^2$.

*TAS characterizations:* The measurements were carried out on the prepared PV devices in a Lakeshore 4-probes cryogenic chamber, where the temperature is controlled by a Lakeshore-360 controller. The frequency-dependent capacitance was characterized with an Agilent B1500 connected to another external capacitance measurement unit. The temperature of the chamber was varied from 220 K to 330 K to record the frequency-dependent capacitance difference. To calculate the built-in voltage and depletion width of the devices, the voltage-dependent capacitance measurements were done with the same instrument.

*Time-resolved PL and PL-QY measurements*: The measurements were performed using a Horiba Jobin Yvon iHR550 Fluorolog system coupled with a Hamamatsu RS5509-73 liquid-nitrogen cooled photomultiplier tube and a calibrated Spectralon-coated Quanta-phi integrating sphere. A 637 nm continuous wave laser (Vortran Stradus) was used as the excitation source for PL measurements. The time-resolved measurements were performed by a pulsed laser (SpectraLED-625) at 637 nm. For extracting correct PL decay curves, the recorded decays were corrected by taking IRF of system into account. The details of the PLQY measurements were described in our previous report[20]. The samples for these measurements were prepared on the glass substrate based on the method described in active layer deposition.

*TA measurement:* The measurements were performed using a mode-lock Ti-saphire oscillator based regenerative amplifier having a center wavelength of 800 nm and pulse width of 45 fs at a repetition rate of 1 kHz. The optical setup for the measurements was a typical pump-probe non-collinear configuration. The output of amplifier was directed into a half-wave plate and a variable neutral density to adjust the pump fluence on the samples. The probe beam (wavelengths ranging 1200 to 1700 nm) was generated by an optical parametric amplifier within almost 1 mJ at 800 nm. A precise motorized stage was used to control the time delay between pump and probe beam. The probe beam was directed on the samples within same excitation area of the pump. The changes in reflection and transmission signals were recorded by a lock-in amplifier. All measurements were carried out at room temperature in solid film formation of QDs on glass substrate within a thickness of ~110-120 nm. The solid films were prepared by the procedure described in active layer deposition on glass substrate.

*ASE and lasing characterization:* The samples were optically pumped by a femtosecond Yt:YAG ORAGAMI laser (NKT Photonics) operating at 1030 nm within a pulse width of 300 fs at a repetition rate of 10 kHz. After passing through a variable neutral density in order to adjust the excitation fluence, the laser beam was focused on to the samples as a stripe shape by using a cylindrical lens. The ASE spectra were collected perpendicular to the excitation utilizing a lens with f = 50 mm and diameter of 2 inch. Using free-space optic, the collected light coupled into Kymera 328i spectrograph (Oxford Instruments. Andor) which equipped by an InGaAs camera, via a f = 200 mm lens through a 100 μm slit. The solid films were prepared by the procedure described in active layer deposition on glass substrate. The thickness of the films is typically ~120 nm.

Lasing spectra were collected perpendicular to the surface of the devices ~20 cm away from the DFB gratings. To do so, we employed a fiber coupled port of the spectrometer using 10-20 μm slit width to achieve high resolution. The integration time of the spectra collection was set to 1 s for all measurements. The picture and video of the DFB lasers were taken by a camera NIT-WiDy-SenS-320V-ST InGaAs camera (Iberoptics Sistemas Ópticos) with an attached SWIR lens. In all measurements, a long pass filter was located at front of the slit to block the pump laser light. Spectral data and threshold analysis of ASE and lasing measurements are presented in Supplementary Note 5.

*DFB grating fabrication:* The gratings were fabricated on pre-cleaned alumina (Ossila Ltd.) substrate in a cleanroom. The PMMA (AR-P 662.04 Allresist) was firstly spin-coated at 4000 r.p.m. for 60 s followed by baking at 150 °C for 2 min. Then, a layer of a conductive polymer (AR-PC 5090.02 Allresist) was spin-coated on top of PMMA layer within a spin speed of 2000 r.p.m., and then baked for 1 min at 90 °C. To pattern the sample, it was transferred to an electron-beam lithography system (Crestec CABL 9000C). Following the lithography, the conductive polymer layer was dissolved in water for 1 min and e-beam resist was developed for 2 min. For etching the alumina substrate, we used reactive-ion etching within 80 sccm of Ar and 20 sccm of $CHF_3$ under 290 V of an RF power for 5 min. Finally, the residual PMMA was then cleaned-off using an oxygen plasma asher and acetone/isopropanol solution cleaning. For DFB laser devices, the PbS/ZnO QDs blend solution was deposited on the pre-patterned alumina substrate through the method described in active layer deposition. The thickness of the active layer is approximately 240 nm (shown in **Figure 4b**). For cross-sectional SEM image, Focused Ion beam (FIB) was employed (Zeiss Auriga), and approximately 100 nm Au layer was deposited on top the DFB laser devices by thermal evaporator (Nano 36 Kurt J. Lesker).

**Supporting Information**

Supporting Information is available from the Wiley Online Library or from the author

**Acknowledgements**

The authors acknowledge financial support from the European Research Council (ERC) under the European Union's Horizon 2020 research and innovation programme (grant agreement no. 725165), the PID2020-112591RB-I00 project funded by MCIN/ AEI /10.13039/501100011033

and the SGR2017AGAUR fund. This project has received funding from the European Union's Horizon 2020 research and innovation programme under the Marie Skłodowska-Curie grant agreement No. 754558. Additionally, this project has received funding from the Spanish State Research Agency, through the "Severo Ochoa" Center of Excellence CEX2019-000910-S, the CERCA Programme / Generalitat de Catalunya and Fundació Mir-Puig. The Authors also acknowledge Prof. J. Climente for fruitful discussions.

**Author contribution**

G.K. co-conceived the idea and supervised the study at all stages. N.T. co-conceived the concept and fabricated, characterized the devices, and analyzed the data. N.T. and G.K. designed the experiments. G.W. fabricated the DFB gratings. A.O. conducted the TA measurements. N.T. analyzed the TA data. M.D. synthesized the PbS QDs. N.T. and Y.W. and G. Kumar conducted TAS measurements. N.T. and Y.W. and analyzed the TAS data. N.T. and S.P. performed time-resolved PL and PL-QY measurements. N.T. and G.K. wrote the manuscript with input from co-authors.

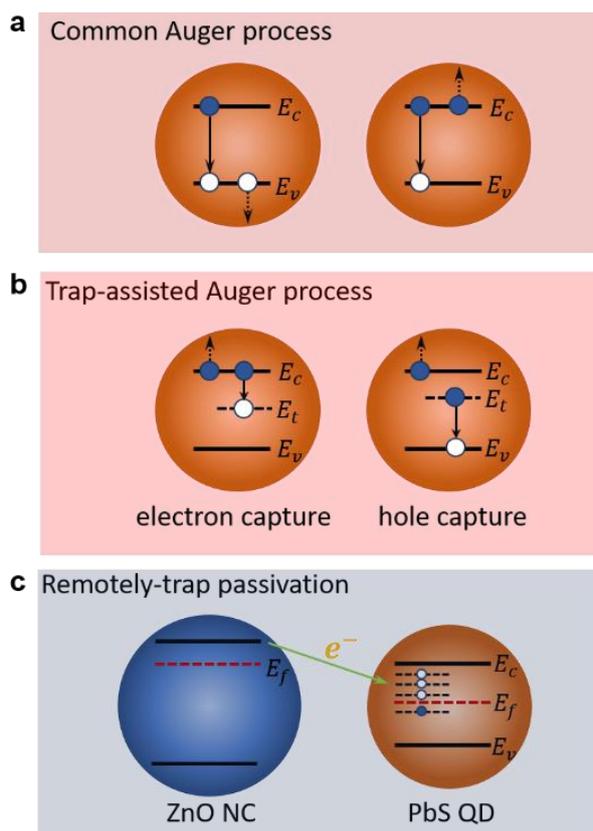

**Figure 1. Concept of suppressing trap-assisted Auger process. a,** Band-to-band Auger recombination, where the recombination and nonradiative energy transfer occurring for band-edge carriers. **b,** trap-assisted Auger mechanism, where the mid-gap trap state captures an electron (hole), and the released energy transferred to the CB electron. **c,** the empty traps (electron acceptor sites) in PbS QD are filled with the electrons of n-type ZnO NC, leading to the reduction of the trap states density.

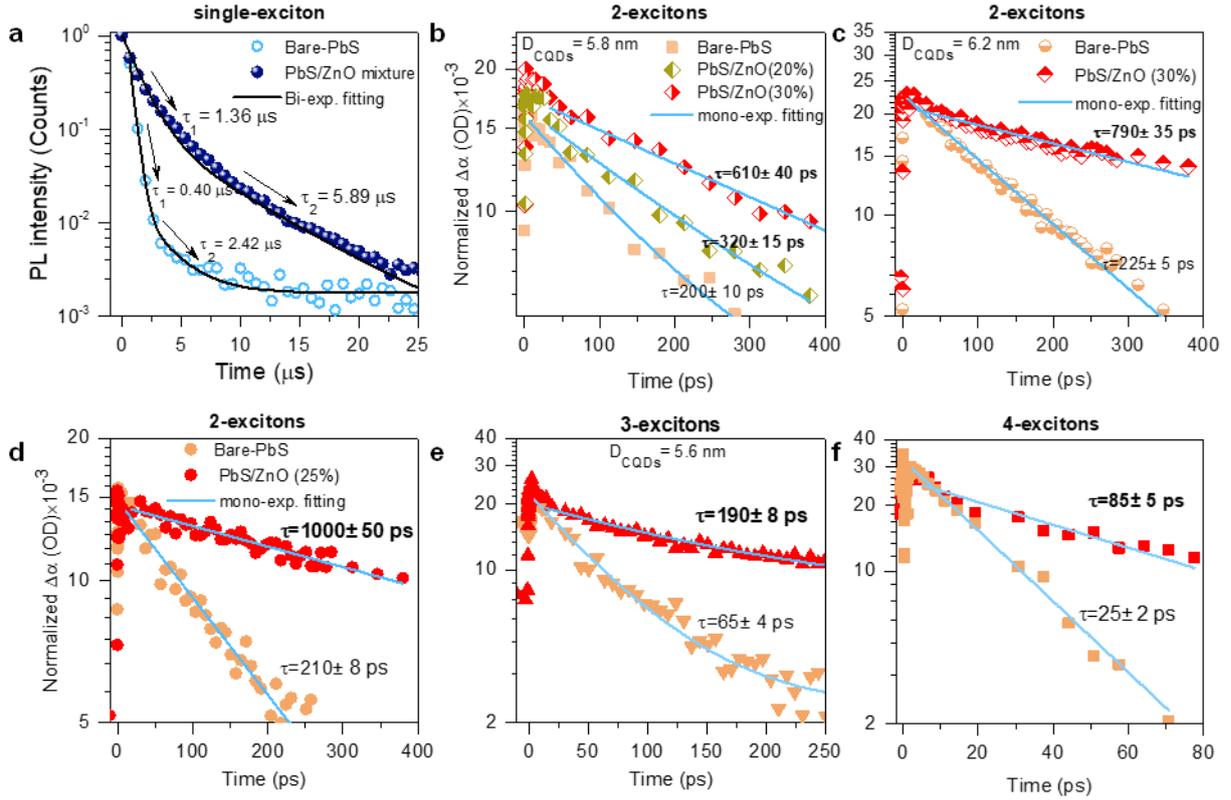

**Figure 2. Single-exciton and multiexciton dynamics. a,** PL decay plots for bare-PbS and Pb/ZnO blend where 30% ZnO NCs mixed with ~5.8 nm PbS-emitter QDs. The decays were recorded at the emission peak wavelength ($\lambda_{emission}$ = 1556 nm) under the pulsed excitation at 637 nm. The decays are fitted with bi-exponential functions (black solid lines) with the component lifetime of $\tau_1$ and $\tau_2$. **b, c,** Dynamics of 2-excitons as a function of ZnO NCs content in binary blend (symbols), obtained from TA spectra of the samples at the band-edge exciton absorption ($1S_e - 1S_h$ transition). "D$_{CQDs}$" denotes the diameter of QDs, D$_{CQDs}$= 5.8 nm **(b)** and D$_{CQDs}$= 6.2 nm **(c).** The blue solid lines represent the corresponding the mono-exponential fitting (lifetime component $\tau$) which extracted from TA measurements based on the method described in ref. 8. **d, e, f,** Kinetics of 2-excitons **(d)**, 3-excitons **(e)** and 4-excitons **(f)** for D$_{CQDs}$= 5.6 nm, where ZnO content is 25% in PbS/ZnO blend sample. The red-symbols represent the PbS/ZnO blend sample, while light-green symbolised for bare-PbS sample. The solid blue lines are the mono-exponential fittings.

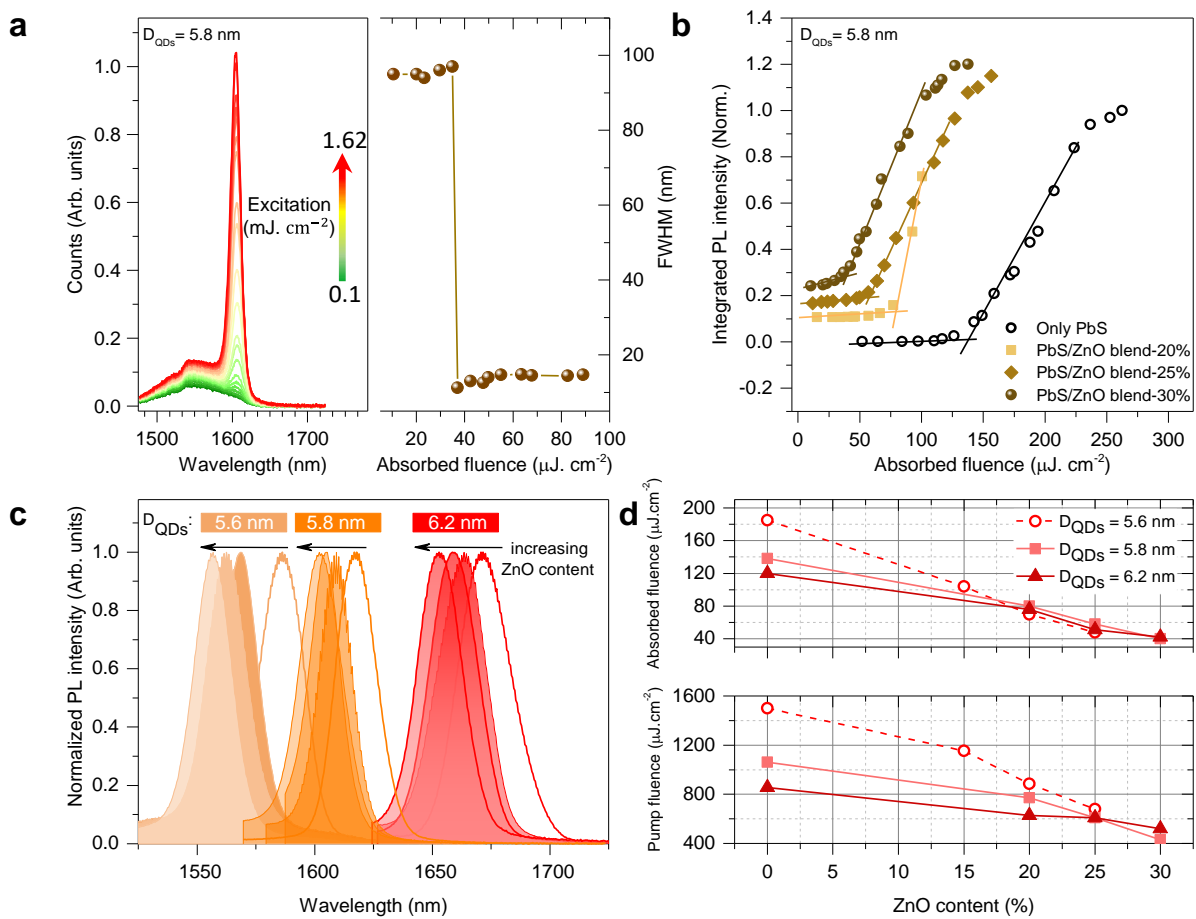

**Figure 3. ASE characterization in bare-PbS emitter and PbS/ZnO binary blend solid films. a,** Pump-dependent ASE spectra of PbS/ZnO (30% loading) binary blend samples, along with corresponding FWHM of each spectrum as function of excitation. "$D_{QDs}$" denotes the diameter of QDs. **b,** Integrated PL spectra depicted as a function of the absorbed fluence at the optical pumping wavelength (1030 nm) for bare-PbS emitter and PbS/ZnO binary blend having ZnO loading of 20%, 25% and 30%. **c,** Collective ASE spectra from a series of bare-PbS QDs, and PbS/ZnO blend samples. Solid lines show the ASE spectra of bare-PbS QDs samples, whilst the colour-filled spectra represent the ASE of binary blend samples with varying ZnO loading as shown by arrows. **d,** The measured ASE thresholds as a function of ZnO content in PbS/ZnO blend sample for three different QDs. The top panel shows the absorbed fluence ASE threshold of the films at 1030 nm (optical pumping wavelength). The bottom panel exhibits the pump fluence ASE threshold.

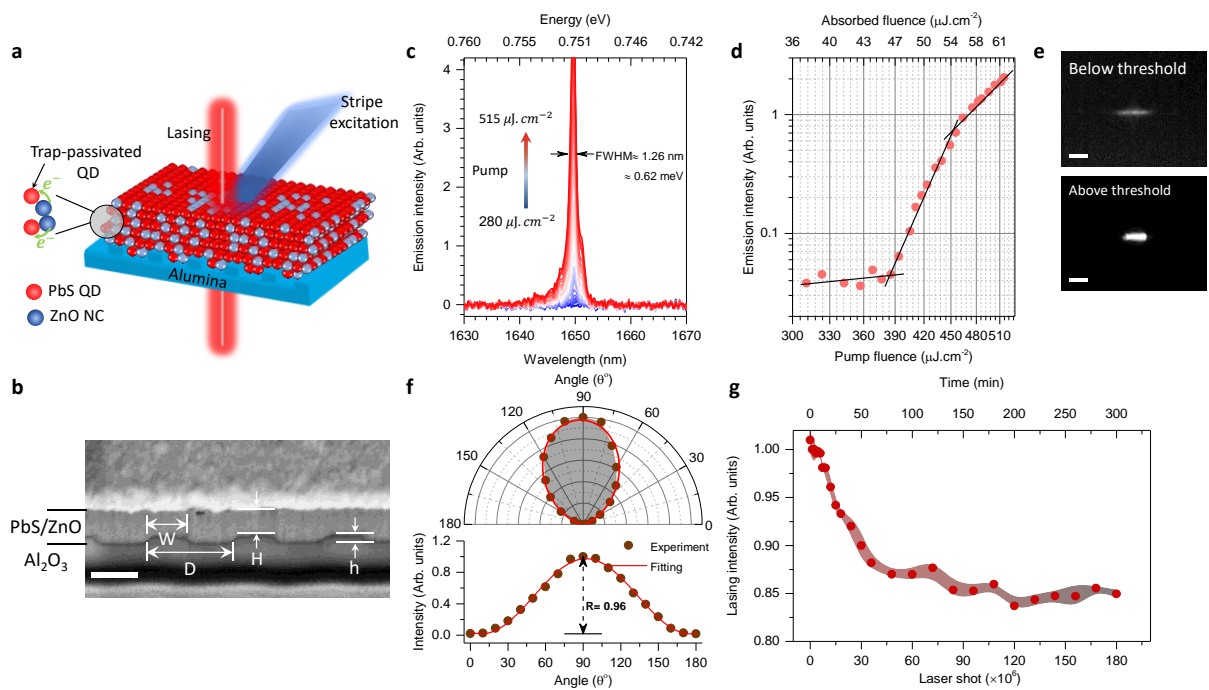

**Figure 4. Lasing performance of a device made of Pb/ZnO binary blend integrated to a second-order grating. a,** Schematic representation of the DFB laser under optical pumping. Red and blue dots symbolized for PbS-emitter QDs and ZnO NCs, respectively. **b,** The SEM image of the PbS/ZnO blend on top of the DFB grating ("H" is the thickness of the PbS/ZnO thin film, width (W), period (D) and height (h) of the grating). The scale bar is 500 nm. **c,** Emission spectra of the DFB laser are plotted as the pump fluence is progressively increased. **d,** Emission intensity versus pump and absorbed fluence (at 1030 nm) of the device (symbols). The solid black lines indicating changes of the slope in log-log scale, as well as showing the lasing threshold at ~385 $\mu$J.cm$^{-2}$. **e,** Infrared picture of the device taken from surface of the device below and above lasing threshold as indicated taken ~20 cm away from the DFB grating. The scale bar is 1 mm. **f,** Normalized emission spectrum of DFB laser above the lasing threshold depicted against of the polarizer angle between the device and the detector. "R" denotes the polarization ratio. **g,** Pump-dependent lasing intensity as a function of the pumping laser shot and corresponding time (symbols). The solid red line indicates the error bar of each symbol, stemming from fluctuation of the lasing intensity. The lasing intensity dropped by 15% after 180 million shots under continuous optical pumping at 10 kHz.

21